\documentclass[sigconf, nonacm]{acmart}

\usepackage{url}
\usepackage[utf8]{inputenc}
\usepackage{xcolor}
\usepackage{fancybox}
\usepackage{graphicx}
\usepackage{amsmath}
\usepackage{amsfonts}
\usepackage{amssymb}
\usepackage{gensymb}
\usepackage{mathtools}
\usepackage{eurosym, booktabs, units}
\usepackage{pgfgantt}
\usepackage{enumitem}
\usepackage{colortbl}
\usepackage{tabularx}
\usepackage{caption}
\usepackage{subcaption}
\usepackage{float}
\usepackage{booktabs}
\usepackage{algorithmicx}
\usepackage{algorithm}
\usepackage{algpseudocode}
\usepackage{siunitx}
\begin{document}

\title{Constrained Neural Ordinary Differential Equations with Stability Guarantees}
\author{J\'an Drgo\v na, Aaron Tuor, Draguna Vrabie}
\affiliation{%
 \institution{Pacific Northwest National Laboratory}
  \city{Richland}
  \state{Washington}
  \postcode{98109}
}

\begin{abstract}

Differential equations  are frequently used in engineering domains, such as modeling and control of industrial systems, where safety and performance guarantees are of paramount importance.
Traditional physics-based modeling approaches require domain expertise
and are often difficult to tune or adapt to new systems.
In this paper, we show how to model discrete ordinary differential equations (ODE) with algebraic nonlinearities  as deep neural networks  with varying degrees of prior knowledge. 
We derive the stability guarantees of the  network layers based on the implicit constraints imposed on the weight's eigenvalues.
Moreover, we show how to use barrier methods to generically handle additional inequality constraints.
We demonstrate  the  prediction accuracy of  learned neural ODEs evaluated on open-loop simulations compared to ground truth  dynamics with bi-linear terms. 
 
\end{abstract}
\maketitle

\section{Introduction}

Ordinary differential equations (ODE)  have numerous applications in various engineering domains such as thermodynamics, mechanics, chemical processes, circuit design, and optimal control. However, the solution of the ODEs  often require sophisticated numerical methods~\cite{Suli_ODE2017}. 
Popular ODE methods are implemented in  Dymola, OpenModelica, MapleSim, Simulink, C, FORTRAN, or Julia and are mostly restricted to users with expert knowledge. 

On the other hand, approaches that bring together deep learning, scientific computing, and differential equations  aim to provide this capability to a broader class of users~\cite{HeZRS15, NIPS2018_7892_NeuralODEs, DeepXDE2019, BattagliaPLRK16, DiffProg2019, RAISSI2019686, Jia2019NeuralJumpStochDE}.
Physics-informed neural networks~\cite{RAISSI2019686}   train fully connected deep neural nets while embedding physics knowledge in the loss function. In this work we take a different approach by directly
specifying the structure of the neural network  to capture the physics, given initial and boundary conditions.
Studies on stability, applying the formal analysis from  dynamical systems to deep learning models derive interesting implications~\cite{IMEXnet2019,NN_Stability2019,NIPS2019_9292,LyapunovNN2018}.
For instance, authors in~\cite{HaberR17} linked the vanishing and exploding gradient problems with the stability of the neural networks interpreted as ODEs and proposed restricted architectures with guaranteed stability. 
We  integrate these findings into our neural ODE framework.

In this paper, we present a novel method for modeling discrete ODE systems as deep neural networks. We demonstrate the possibility of incorporating varying degrees of prior knowledge combining physics-based and purely data-driven modeling in a unified framework. Moreover, we show how to impose stability guarantees and inequality constraints on the layers of arbitrary neural architectures. We apply the proposed method to the identification of a ODE model with bi-linear terms simulating a thermal system. We show that embedding constraints and stability regularizations can provide advantages in sample efficiency,  generalization, as well as physically consistent trajectories.

\section{Methods}

Section~\ref{sec:methods:ODE}  describes the architecture of discrete neural  ODE with possible variations and extensions.
In section~\ref{sec:methods:ODE}, we derive the stability guarantees of generic neural architectures by constraining eigenvalues of the layer weights. Moreover, we introduce a generic method for imposing inequality constraints on hidden states of deep neural networks. The proposed method for modeling of neural ODE systems is demonstrated in section~\ref{sec:casestudy} on a  case study from the energy domain.

\subsection{Neural Ordinary Differential Equations}
\label{sec:methods:ODE}

\paragraph{Ground Truth ODE:}
Our task is to model the dynamics of an unknown ground truth ODE system in discrete-time  with linear dynamics and  bi-linear algebraic form:
\begin{subequations}
\label{eq:truth}
\begin{align}
  {\bf x}_{k+1} & = {\bf A} {\bf x}_k + {\bf B } {\bf u}_k + {\bf E} {\bf d}_k ,   \label{eq:truth:x} \\
   {\bf u}_k  & = {\bf a}_k {\bf H} {\bf b}_k + {\bf h},   \label{eq:truth:u} 
\end{align}
\end{subequations}
where ${\bf x}_k \in \mathbb{R}^{n_x}$ is the system state,
${\bf y}_k \in \mathbb{R}^{n_y}$ is the system output, 
${\bf u_k} \in \mathbb{R}^{n_u}$ is the algebraic input, and ${\bf d}_k \in \mathbb{R}^{n_d}$ is measured disturbance at time $k$. 
The bi-linearity is defined via algebraic~\eqref{eq:truth:u} with  linear terms ${\bf H}$, ${\bf h}$ and  inputs ${\bf a}_k \in \mathbb{R}^{n_a}$ and ${\bf b}_k \in \mathbb{R}^{n_b}$.

\paragraph{Discrete Neural ODE:}
Single time step of our neural ODE model has the following form:
\begin{subequations}
\label{eq:sysID:ODE}
\begin{align}
    f_{\texttt{ODE}}({\bf x}, {\bf a}, {\bf b}, {\bf d})  &  =  f_{\texttt{SSM}}({\bf \tilde{A}} {\bf x} + {\bf \tilde{ B} } {\bf u} + {\bf \tilde{E}} {\bf d}) \label{eq:sysID:ODE:lin} \\
    {\bf u} &  = h_\Theta({\bf a}, {\bf b}) \label{eq:sysID:ODE:bilin}
\end{align}
\end{subequations}
By stacking multiple layers of~\eqref{eq:sysID:ODE} with shared weights, we can construct time-invariant ODE model with arbitrary depth $N$, where each layer corresponds to the one-time step defined by the  sampling time  of the training data.
The main dynamics~\eqref{eq:sysID:ODE:lin} is given as a state space model (SSM)  with parameters ${\bf \tilde{A}}$, ${\bf \tilde{B}}$, ${\bf \tilde{E}}$, and activation function $f_{\texttt{SSM}}$.
In this paper, the  $f_{\texttt{SSM}}$
is given as an identity operation to model linear dynamics. 
We further differentiate the  baseline neural ODE model  into  three forms  with  varying degrees of prior knowledge about the   algebraic~\eqref{eq:sysID:ODE:bilin}.
In case of no prior knowledge, we  use a  \textit{black-box}  ODE  ($\texttt{ODE}_\texttt{B}$), where~\eqref{eq:sysID:ODE:bilin}  can be modeled by a standard multi-layer fully connected neural network.   
For the purposes of this paper, we have chosen two layers  with  $\texttt{ReLU}$ activations and $8$ hidden units.
In practice, we may often know the structure of the underlying algebraic~\eqref{eq:sysID:ODE:bilin},  for instance, based on  known  physical laws governing the system dynamics.
In our case, this equation is given as bi-linear term ${\bf u}   = {\bf a} {\bf \tilde{H}} {\bf b} + {\bf \tilde{h}}$, with learnable parameters $\Theta = \{ {\bf \tilde{H}}, {\bf \tilde{h}} \}$.
We will refer to such a model as a \textit{gray-box}  ODE  ($\texttt{ODE}_\texttt{G}$).
When the structure, as well as the parameters of~\eqref{eq:sysID:ODE:bilin} are known,  e.g., obtained from the engineering sheets,
we use the \textit{white-box} ODE ($\texttt{ODE}_\texttt{W}$)
with  given constants ${\bf {H}}$, ${\bf {h}}$ of its  bi-linear term. 

 \paragraph{Variations and Extensions:} To model the time-varying dynamics, we can stop sharing the weights in the successive layers to generate piecewise-linear approximations.
Another  extension is to  use the nonlinear activation function $f_{\texttt{SSM}}$ in~\eqref{eq:sysID:ODE:lin}, and  increasing the  depth of a single time step model.
Similarly, it is straightforward to extend the input space  by state variables in~\eqref{eq:sysID:ODE:bilin} for  approximating differential algebraic equations (DAE)~\cite{AMODIO1997135}.  Moreover, we can structurally prior arbitrary algebraic terms, given as polynomials constructed by stacking multiple bi-linear terms. 
An extensive list of structural priors and possible applications is beyond the scope of this paper.

\subsection{Optimization with Stability Guarantees and Constraints Handling}
\label{sec:methods:constr}

\paragraph{Eigenvalues of the Layer Weights:}
The Perron–Frobenius theorem~\cite{Knill:LinAlg:11} states that the
row-wise maximum and minimum of nonnegative square matrix ${\bf {A}}$  defines the  upper and lower bound of its dominant eigenvalue. 
We use this to constrain the  eigenvalues of the  weights ${\bf \tilde{A}}$ to enforce the stability of the layer forward pass.
This constraint is formulated as:
  \begin{equation}
  {\bf M} = 1 - 0.1 \sigma({\bf M}')
    \label{eqn:1}
  \end{equation}
  \begin{equation}
  {\bf \tilde{A}}_{i, j} =  \frac{\exp({\bf \tilde{A}'}_{ij}) {\bf M }_{i,j}}{\sum_{k=1}^{n_x} \exp ({\bf \tilde{A}}'_{ik})}
    \label{eq:eigenvalue_A}
  \end{equation}
 Where the matrix ${\bf M}$ is modeling  damping given as a function of parameter  ${\bf M}' \in \mathbb{R}^{n_x\times n_x}$.  
We use softmax  regularized rows of the ${\bf \tilde{A}}'$ matrix
in elementwise multiplication with ${\bf M}$ to generate the new weight matrix  ${\bf \tilde{A}}$ of the state dynamics  used in \eqref{eq:sysID:ODE:lin}.
With dominant eigenvalue to be  less or equal to one, the stability of the learned dynamics of the discrete system is guaranteed.
Additionally, by having the eigenvalues of layer weights  close to one for discrete time,  or zero for continuous time systems, respectively, the well-posedness of the learning problem is estabilished by preventing exploding and vanishing gradients~\cite{HaberR17}.

\paragraph{Inequality constraints via penalty method:}
For handling the inequality constraints we employ the penalty method for constrained optimization~\cite{BenNem:opt:01,BoyVan:ConOpt:04}.
The principle idea is based on penalizing the constraints in the objective of the unconstrained optimization problem. 
In particular, we use  ReLU units to model the violations of  inequality constraints:
\begin{subequations}
\label{eq:ReLU_ineq}
    \begin{align}
    \underline{{\bf x}}_k   \leq {{\bf x}_k} +  {{\bf s}_k^{\underline{x}}} \ \ & \cong \ \ {\bf s}_k^{\underline{x}}  = \texttt{ReLU}(-{\bf x}_k+\underline{{\bf x}}_k) \\
  {{\bf x}_k} - {{\bf s}_k^{\overline{x}}} \leq \overline{{\bf x}}_k  \ \ & \cong \ \ {\bf s}_k^{\overline{x}}  = \texttt{ReLU}({\bf x}_k-\overline{{\bf x}}_k)
 \end{align}
 \end{subequations}
 Here,  ${\bf s}_k^{{x}} = {\bf s}_k^{\underline{x}} + {\bf s}_k^{\overline{x}}$ define  joint slack variables representing the magnitude of the constraints violation, and 
 $\underline{{\bf x}}_k$ and $\overline{{\bf x}}_k$ stand for  time-varying lower and upper bound on the variable ${{\bf x}_k}$, respectively. Analogically the constraints can be defined for all model variables. 
In this paper, we impose the inequality constraints on states 
 ${{\bf x}}_k$ and algebraic inputs ${{\bf u}}_k$, to keep their trajectories within physically realistic bounds. We refer to the constrained ODE models as $\texttt{cODE}$.
 The proposed method for constraints handling is generic and not limited to any specific neural architecture.
 The constraints can be imposed on model outputs, hidden states, or their derivatives.  

\paragraph{Loss function:}
The objective  penalizes the  deviations of the model response defined by~\eqref{eq:sysID:ODE} from the training data  obtained from simulating the ground truth system~\eqref{eq:truth}  over $N$ time steps  generating sequences of vectors, $\mathcal{X} = {\bf x}_0, ...{\bf x}_N$, $\mathcal{A} = {\bf a}_0, ... {\bf a}_N$, $\mathcal{B} = {\bf b}_0, ... {\bf b}_N$, $\mathcal{D} = {\bf d}_0, ...{\bf d}_N$. 
 We assume that for optimization, we only have access to one observable variable ${\bf x}_{k,i}$  denoted by index $i$.
The model is given only an initial state ${\bf x}_0$, together with system inputs  $\mathcal{A}$ and $\mathcal{B}$, and disturbance $\mathcal{D}$ trajectories to  produce sequences of state predictions, $\tilde{\mathcal{X}} = \tilde{{\bf x}}_0, ...\tilde{{\bf x}}_N$, as well as slack variables referring to the state and hidden inputs constraints violations ${{S}^{x}} = {{\bf s}}_0^{x}, ...{{\bf s}}_N^{x}$ and ${\mathcal{S}^{u}} = {{\bf s}}_0^{u}, ...{{\bf s}}_N^{u}$, respectively.
Constraints violations are penalized in the objective with weighting factors $\lambda$ and $\mu$.
 The multi-objective Mean Squared Error (MSE) loss over given $N$-step prediction horizon is then:
\begin{equation} \label{eqn:nstepmse}
\mathcal{L}_{\texttt{MSE}}(\tilde{\mathcal{X}}, \mathcal{X} | \tilde{\bf A}, \tilde{\bf B}, \tilde{\bf E}, \Theta) = \frac{1}{N}  \sum_{k=1}^N  \big(  ||{\bf \tilde{x}}_{k, i} - {\bf x}_{k, i}||^2_2  + \lambda||{\bf s}_{k}^x||^2_2  + \mu||{\bf s}_{k}^u||^2_2 \big)
\end{equation}
In the general neural ODE model given by~\eqref{eq:sysID:ODE} we optimize the 
$\tilde{\bf A}$, $\tilde{\bf B}$, $\tilde{\bf E}$ parameter matrices
of the linear dynamics. Moreover, in the case of $\texttt{ODE}_\texttt{B}$ and $\texttt{ODE}_\texttt{G}$, also the  $\Theta$ parameters of the approximated algebraic~\eqref{eq:sysID:ODE:bilin} are optimized.
The models are implemented\footnote{Code for reproducing our experiments is available at:\\ \url{https://github.com/pnnl/neural_ODE_ICLR2020}} in Pytorch~\cite{paszke2019pytorch}. 

\section{Numerical Case Study}
\label{sec:casestudy}
We compare empirical results for identification of models with and without constraints ($\texttt{cODE}$ vs $\texttt{ODE}$), and with three degrees of prior knowledge about the algebraic interaction between input variables ($\texttt{ODE}_\texttt{B}$, $\texttt{ODE}_\texttt{G}$, $\texttt{ODE}_\texttt{W}$). We simulated the true dynamics (\eqref{eq:truth}) of a simple building thermal system with state ${\bf x} \in \mathbb{R}^4$ whose elements correspond to wall ($x_1$), ceiling ($x_2$), floor ($x_3$), and ambient room temperature ($x_4$, the observed state). The bi-linear term is a heat flow equation with constant parameter of specific heat capacity  ${\bf H} =  c_p, {\bf h} = 0$, and two variables, mass flow ${\bf a}_{k} =  \dot{{\bf m}}_{k}$, and difference of supply and return temperature ${\bf b}_{k} = \Delta {\bf T}_{k}$.
 The corresponding control input signals  $\mathcal{A}$, and $\mathcal{B}$ are generated as a  sine and cosine wave, respectively, 
 with the period of one day and amplitudes as nominal physical values of the true model. 
 The  disturbance signals $\mathcal{D}$ represent the historical environmental conditions. We use the $2^{nd}$, $3^{rd}$, and $4^{th}$ weeks of simulation as train, validation, and test sets (each containing 2016 contiguous time-steps).
 We chose this limited time period to demonstrate generalization capability with limited training samples.

\begin{figure}
\parbox{.48\textwidth}{
    \centering
    \includegraphics[width=.48\textwidth]{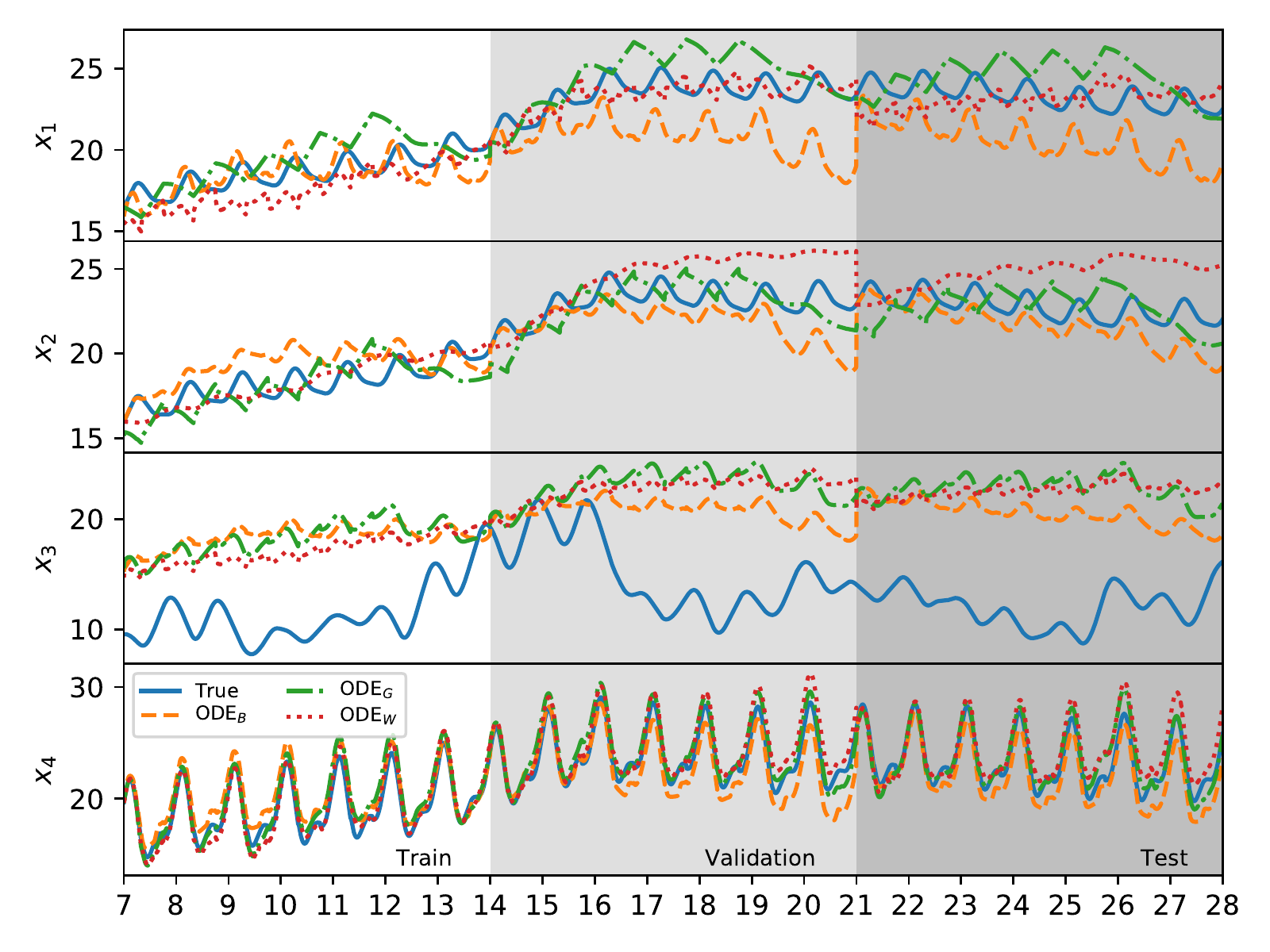}
    \caption{Non-constrained model trace.}
    \label{fig:traj_noconstr}
 }
 \hfill
 \parbox{.48\textwidth}{
    \centering
    \includegraphics[width=.48\textwidth]{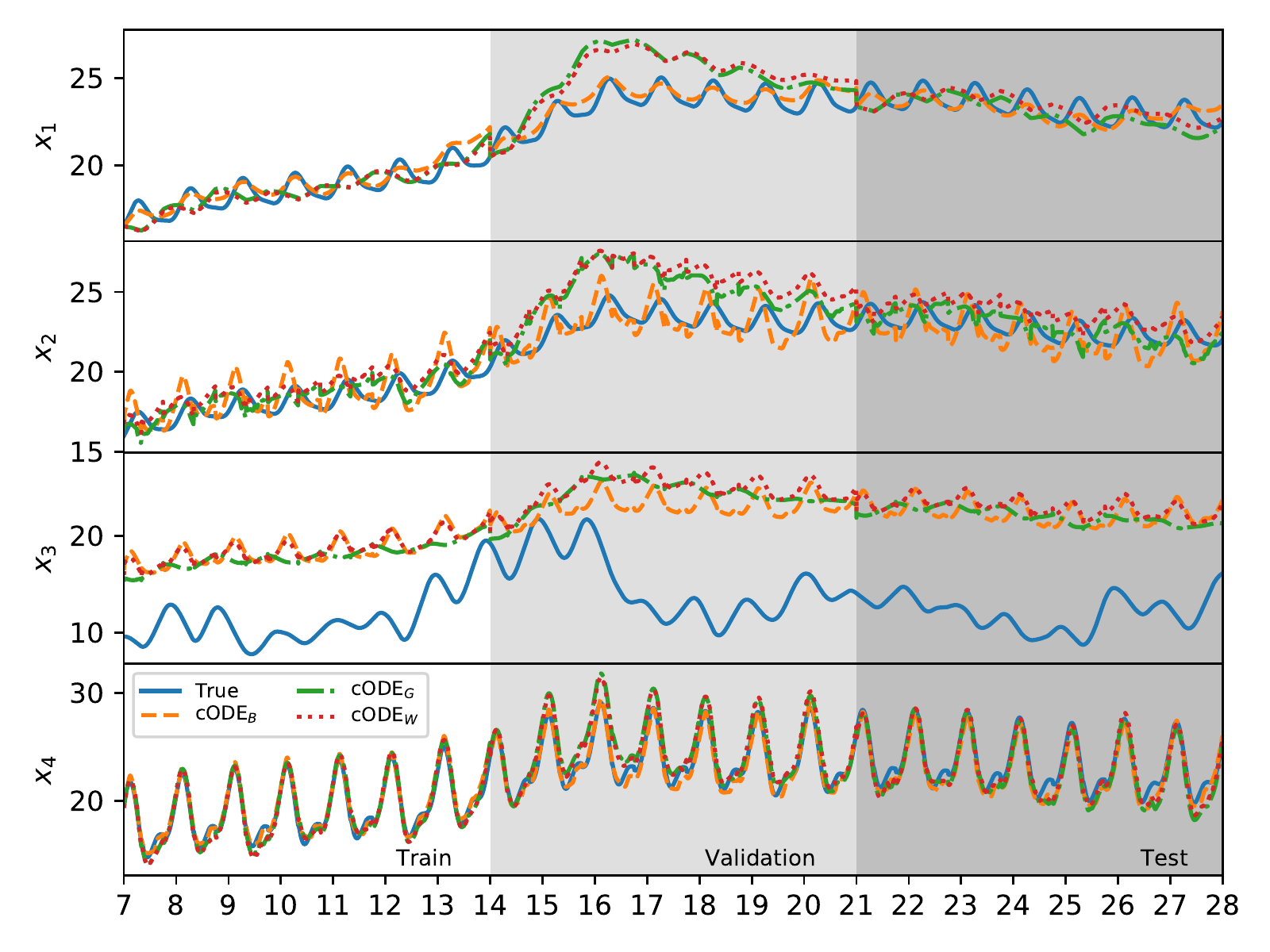}
    \caption{Constrained model trace.}
    \label{fig:traj_constr}
    }
\end{figure}

For black, gray, and white-box ODE models with and without state and control restraints as described in section \ref{sec:methods:ODE} we train models with N-step prediction objective, $N \in \{2^3, ...,2^7\}$. For each of these 30 (model, N-step objective) pairs we train 30 models from random parameter initializations with full batch AdamW~\cite{loshchilov2017decoupled} updates (step-size ranging from 0.001 to 0.03) for 15,000 epochs. 
We evaluate the prediction performance of the learned models for both the N-step prediction training objective, and open-loop MSE for model simulation over the test set with $T=2016$ time steps: $ \frac{1}{T} \sum_{t=1}^T ({\bf x}_{k, 4}^{\text{ID}} - {\bf \tilde{x}}_{k, 4})^2$.
Tables \ref{tab:nstepmse}, and \ref{tab:openmse} show the $N$-step and open-loop MSE respectively on the test set for each $N$-step prediction horizon training objective. As $N$ increases, models tend to higher $N$-step MSE, and lower open-loop MSE. This makes sense as, while training with a longer prediction horizon is a more difficult learning objective, the longer horizons provide closer approximations to the open-loop behavior of the learned models. Black and gray-box models without constraints fail to realize gains from the longest horizon, whereas the best performing models were constrained models with a 128-step prediction horizon objective. Interestingly, the best performing model was a $\texttt{cODE}_\texttt{B}$, suggesting that given physically reasonable constraints, acceptable dynamics models can be learned given less prior knowledge of the true system.
Figures \ref{fig:traj_noconstr} and \ref{fig:traj_constr} show open-loop simulations from non-constrained and constrained models, respectively. The solid blue line indicates the true system trajectory. Unconstrained models' trajectories drift more over time, especially for the unobserved variables. Notably, the $\texttt{cODE}_\texttt{B}$ model does a remarkable job of tracking all state variables, excepting $x_3$ (floor temperature), which has the weakest connection with the observed $x_4$.  

\begin{table}
\centering
\begin{tabular}{lllllll}
\toprule
{$N$} &   8   &   16  &   32  &   64  &   128\\
\midrule
$\texttt{ODE}_\texttt{B}$    & 0.04 & 0.13 & 0.47 & 0.31 & 0.41  \\
$\texttt{ODE}_\texttt{G}$    & 0.08 & 0.33 & 0.92 & 0.82 & 0.65 \\
$\texttt{ODE}_\texttt{W}$   & 0.08 & 0.31 & 0.92 & 0.81 & 0.65 \\
$\texttt{cODE}_\texttt{B}$ & 0.03 & 0.13 & 0.46 & 0.30 & 0.35 \\
$\texttt{cODE}_\texttt{G}$  & 0.08 & 0.33 & 0.92 & 0.91 & 0.58 \\
$\texttt{cODE}_\texttt{W}$ & 0.08 & 0.33 & 0.92 & 0.82 & 0.59  \\
\bottomrule
\end{tabular}
\caption{Best N-step prediction MSE.}
\label{tab:nstepmse}
\end{table}

\begin{table}
\centering
\begin{tabular}{lllllll}
\toprule
{$N$} &    8   &    16  &   32  &   64  &   128\\
\midrule
$\texttt{ODE}_\texttt{B}$    &  9.93 &  3.75 & 5.15 & \textbf{2.24} & 2.59  \\
$\texttt{ODE}_\texttt{G}$     & 19.0 & 23.0 & 4.19 & \textbf{0.91} & 2.56  \\
$\texttt{ODE}_\texttt{W}$    & 19.6 & 19.2 & 6.58 & 5.24 & \textbf{3.81} \\
$\texttt{cODE}_\texttt{B}$ &  3.44 &  3.48 & 4.94 & 2.47 & \textcolor{red}{\textbf{0.22}}  \\
$\texttt{cODE}_\texttt{G}$  & 19.5 & 19.5 & 4.68 & 2.96 & \textbf{0.56} \\
$\texttt{cODE}_\texttt{W}$ & 19.9 & 19.6 & 6.91 & 7.60 & \textbf{0.41} \\
\bottomrule
\end{tabular}
\caption{Best open-loop prediction MSE.}
\label{tab:openmse} 
\end{table}

\section{Conclusions and Future Work}
This work presents novel methods for modeling discrete ordinary differential equations (ODE) as neural networks with i) stability guarantees based on eigenvalue regularization of the layer weights and ii) time-varying inequality constraints. Both i) and ii) are implemented using standard operations available in popular deep learning libraries.  We demonstrate remarkable generalization and the ability to learn physically consistent ODE dynamics from a limited amount of training data. Empirical results underscore the  advantages of using penalty methods for minimizing the constraints violation in the problem's loss function. Thus, enabling model safety assessment and certification. 

The future work includes applying the neural ODEs and DAEs to model large-scale physical systems with various types of dynamic and algebraic nonlinearities. Computational efficiency and scalability of the proposed neural ODE can be further compared with classical ODE solution methods. 
For practical purposes, the authors intend to develop a library of physics-informed ODE and DAE priors commonly occurring in various engineering domains for user-friendly \textit{gray-box} modeling.
The authors also intend to explore the use of neural ODEs in
model-based deep learning approaches to constrained optimal control for physical systems.
The  convergence guarantees can be  obtained by means of Lyapunov stability analysis of the loss function.
Another open research avenue is the development of customized optimizers for the solution of constrained optimization problems.
Finally, the generic nature of the proposed methods for enforcing stability and constraints handling can be explored on various neural architectures and learning tasks.

\subsubsection*{Acknowledgements}
This work was funded by the Mathematics for Artificial Reasoning in Science investment at the Pacific Northwest National Laboratory (PNNL). 

\bibliographystyle{acm}
\bibliography{bib}

\section{Appendix}
In this appendix we present additional visualizations comparing model performance, eigenvalues and heatmaps for learned transition matrices, and a comparison with a preliminary Physics-Informed Recurrent Neural Network Model (PI-RNN) without  constraints on the principal learned dynamics matrix ${\bf \tilde{A}}$ or its hidden states.

\subsection{Additional performance visualizations}
\begin{figure}[htb]
 \parbox{.48\textwidth}{
    \centering
    \includegraphics[width=.48\textwidth]{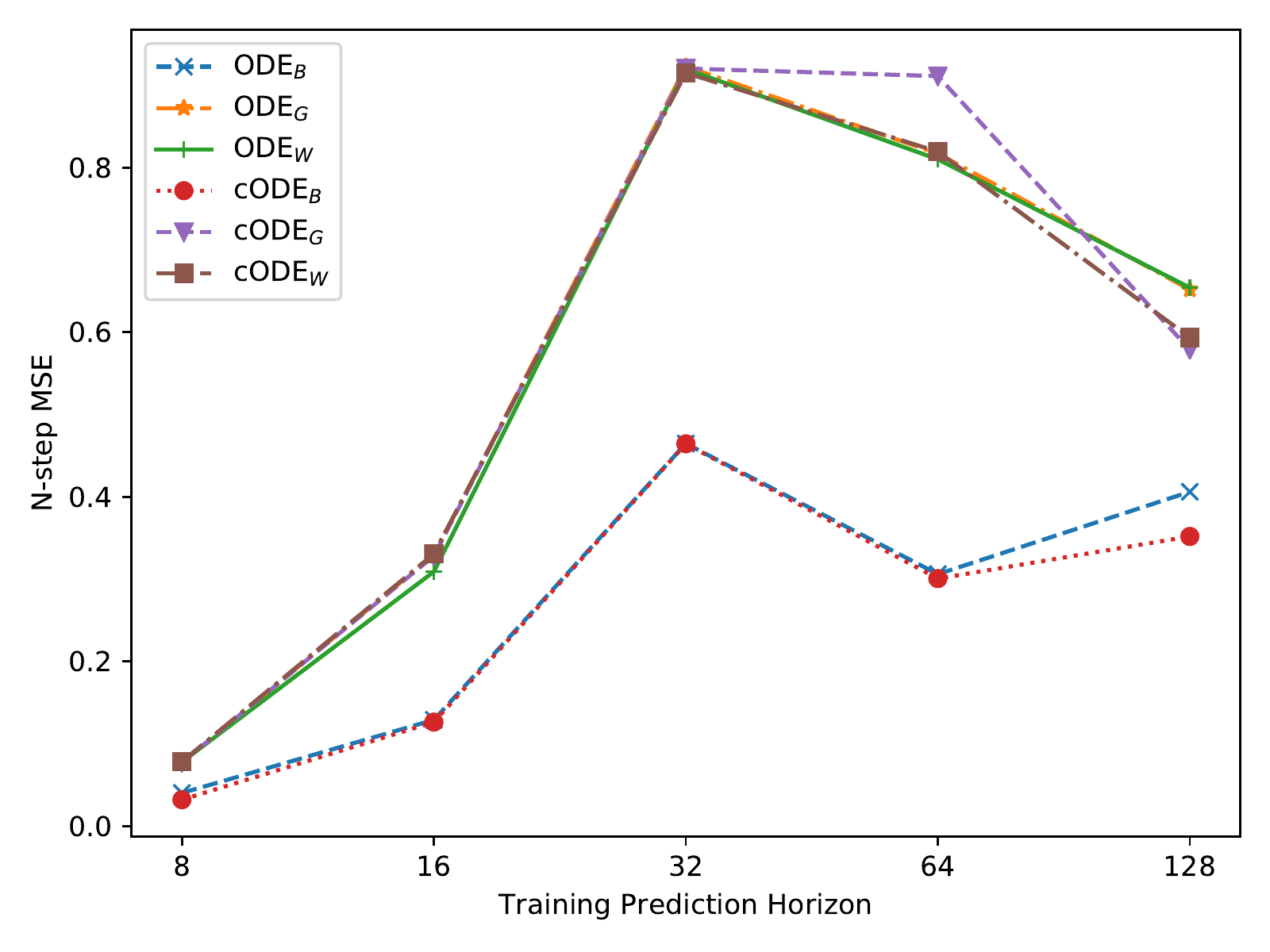}
    \caption{N-step best MSE}
    \label{fig:NstepMSE}
    }
     \hfill
    \parbox{.48\textwidth}{
    \centering
    \includegraphics[width=.48\textwidth]{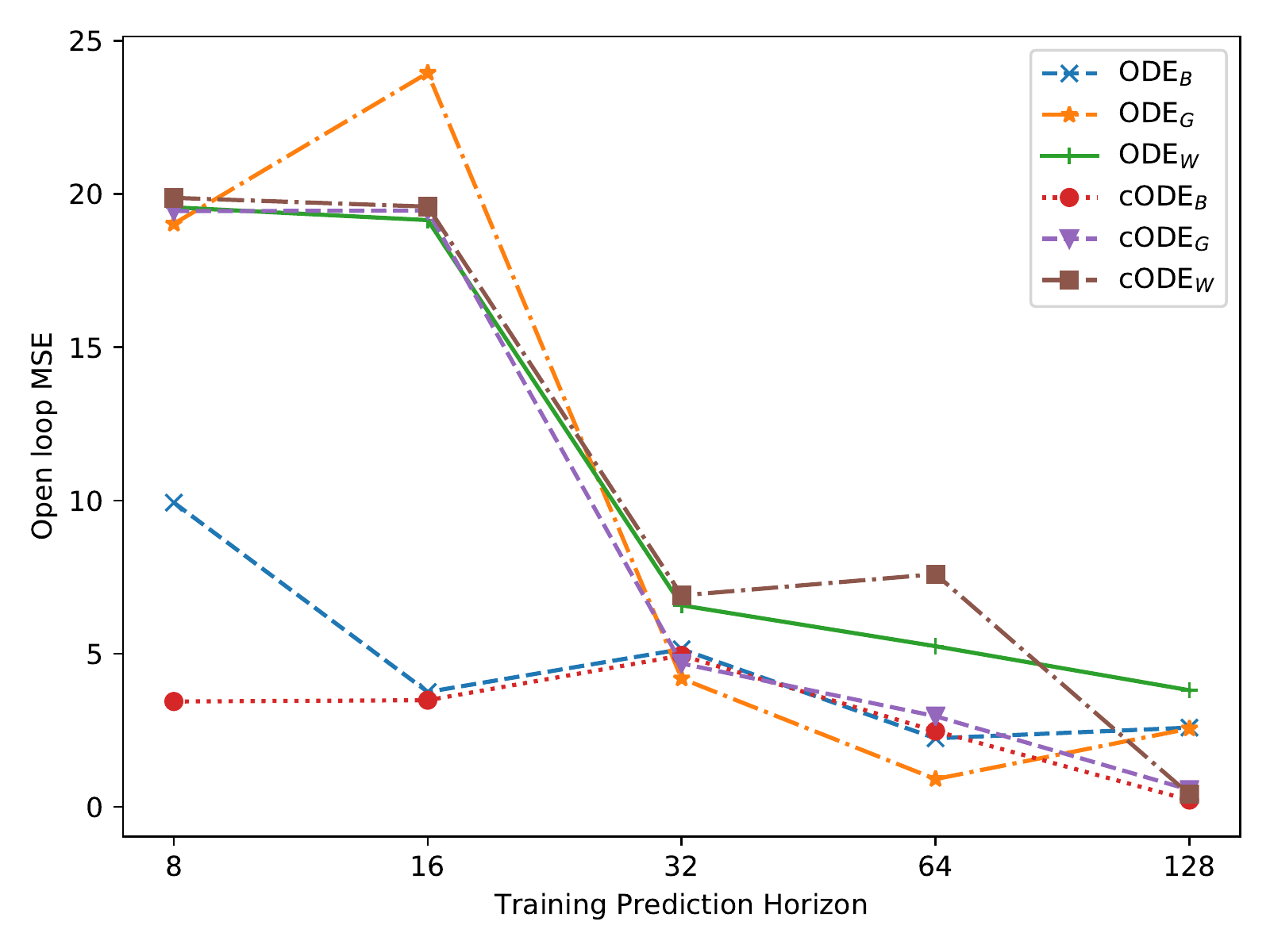}
    \caption{Open-loop best MSE}
    \label{fig:OL_MSE}
 }
\end{figure}

Figures~\ref{fig:NstepMSE}  and~\ref{fig:OL_MSE}  visualize the influence of the increasing prediction horizon $N$ on the open loop MSE and $N$-step MSE loss, reported in Tables~\ref{tab:nstepmse}  and~\ref{tab:openmse} , respectively. The increasing trend of the $N$-step MSE with larger prediction horizon $N$ is given by the increasing complexity of the learning problem,
which is correlated with the increased accuracy of the learned models in open-loop simulations.

\subsection{Ground Truth Physical Model}

The system \eqref{eq:truth} represents a simple  building thermal system with states ${\bf x}_k \in \mathbb{R}^4$ whose elements are  wall (${\bf x}_{k,1}$), ceiling (${\bf x}_{k,2}$), floor (${\bf x}_{k,3}$), and  room temperature (${\bf x}_{k,4}$, the observed state).  Control inputs \eqref{eq:truth:u} represent  heat flow equation of the building's radiator ${\bf u}_{k} = \dot{{\bf m}}_{k} c_p \Delta {\bf T}_{k}$, with mass flow $\dot{{\bf m}}_{k}$, specific heat capacity $c_p$, and temperature difference of the emission system $\Delta {\bf T}_{k}$.   The 
disturbances  ${{\bf d}_k \in \mathbb{R}^3}$ represent ambient temperature (${\bf d}_{k,1}$), solar irradiation (${\bf d}_{k,2}$), and internal heat gains (${\bf d}_{k,3}$), respectively.
We generate  the state trajectories $\mathcal{X}$  for the system identification by simulating the   model \eqref{eq:truth}  with initial conditions of ${\bf x}_0 =$
 \SI{20}{\celsius}, given the measured disturbance trajectories  ${\mathcal{D}}$.
In practice,  ${\mathcal{D}}$ is obtained from  weather forecast.

An interesting property of the thermal models of the buildings is that their transition matrix $\tilde{{\bf A}}$ is, in general, non-negative with stable eigenvalues.
This feature motivates the use of the eigenvalue regularization given by  \eqref{eq:eigenvalue_A}. In this case, the damping factor ${\bf M}$ can be physically interpreted as heat losses of the building envelope. Hence physical insights can be used for tuning of the proposed model to different building types.
The penalty constraints on the state trajectories are derived based on the physically meaningful values for the building.

\subsection{Structured RNN Model}
Here we introduce a preliminary model with neither eigenvalue constraints on the learned ${\bf \tilde{A}}$ matrix nor barrier penalties in the learning objective. Considering the same limited knowledge of the underlying dynamics as in the $\texttt{ODE}_B$ model we introduce a preliminary model $\texttt{S-RNN}$. The discrete $\texttt{SSM}$ model remains the same as equation \ref{eq:sysID:ODE:lin}, but with a two layer neural network modeling the underlying bi-linear algebraic term as follows: 
\begin{subequations}
\label{eq:sysID:rnn}
\begin{align}
    f_{\texttt{S-RNN}}({\bf x}, {\bf a}, {\bf b}, {\bf d})  &  =  f_{\texttt{SSM}}({\bf \tilde{A}} {\bf x} + {\bf \tilde{ B} } {\bf u} + {\bf \tilde{E}} {\bf d}) \label{eq:sysID:PIRNN:lin} \\
    {\bf u} &  = \texttt{ReLU}\bigl({\bf \tilde{W}_2} {\bf h_1} + {\bf \tilde{W}_3} \begin{bmatrix}{\bf a} \\{\bf b}\end{bmatrix}\bigr)\\
    {\bf h_1} & = \texttt{ReLU}\bigl({\bf \tilde{W}_1}\begin{bmatrix}{\bf a} \\{\bf b}\end{bmatrix}\bigr)
    \label{eq:rnnbilin}
\end{align}
\end{subequations}
where the matrices ${\bf \tilde{W}}$ are additional learned parameters for the algebraic equation approximation via neural network.

\begin{figure}
\parbox{.48\textwidth}{
    \centering
    \includegraphics[width=.48\textwidth]{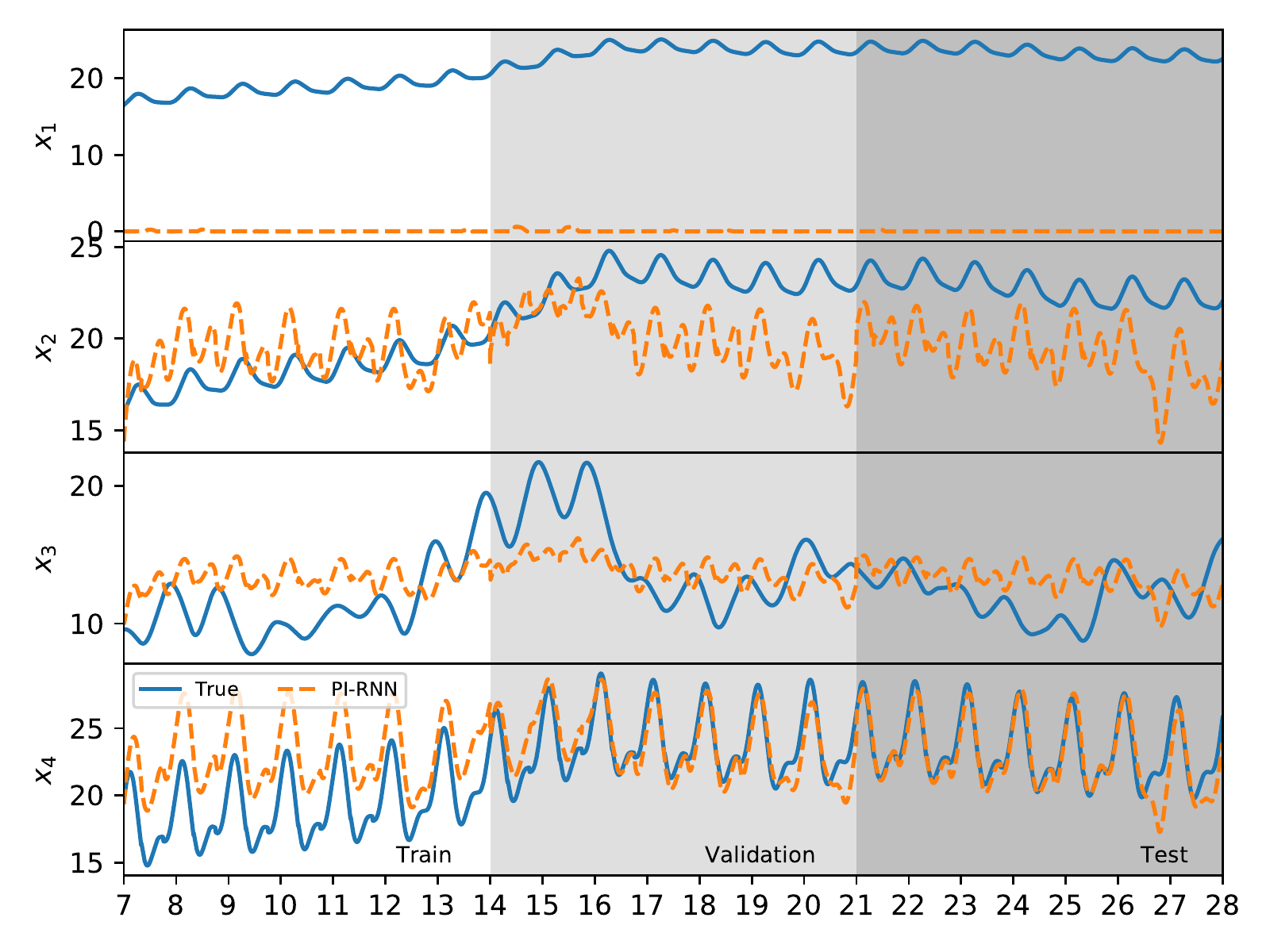}
    \caption{Open-loop trajectory for best performing S-RNN model.}
    \label{fig:rnntrace}
 }
 \hfill
 \parbox{.48\textwidth}{
    \centering
     \begin{tabular}{llllll}
\toprule
{} &    8   &    16  &   32  &   64  &   128  \\
\midrule
Open-loop       &  2.03 &  1.22 & 1.13 & 3.10 & 3.04 \\
$N$-step        & 0.02 & 0.09 & 0.20 & 1.02 & 0.64\\
\bottomrule
\end{tabular}
    \captionof{table}{Structured RNN MSE for $N$-step and open-loop prediction.}
    \label{tab:rnn}
    }
\end{figure}

Figure~\ref{fig:rnntrace} shows the open-loop trace of trained $\texttt{S-RNN}$ on training, validation, and test set, respectively. The trajectories can be directly compared with Figures~\ref{fig:traj_noconstr} and~\ref{fig:traj_constr}, at first glance it is visible that although capable of accurate prediction and generalization of the observed state, the $\texttt{S-RNN}$ fails to capture the dynamics of the unobserved states, in contrast with $\texttt{ODE}$ models. 
The corresponding  open-loop MSE and $N$-step MSE with increasing values of the model prediction horizon $N$ are given in Table~\ref{tab:rnn}. We observe that in contrast with $\texttt{ODE}$ models $\texttt{S-RNN}$ fails to leverage the advantage of the larger prediction horizons to improve its accuracy.

\subsection{Effect of the Learned Eigenvalues}
Table~\ref{tab:eig} compares the eigenvalues of the learned transition matrix $\tilde{{\bf A}}$ with the eigenvalues of the ground truth model.
All physics-informed models accurately learned the stable dominant eigenvalue of the primary dynamics.  However, differences arise when comparing the rest of the eigenvalue spectrum.
The eigenvalues of the trained $\texttt{ODE}$   models have, in general, shorter Euclidean distance from the ground truth values compared to $\texttt{S-RNN}$ model.
However, the eigenvalues of the constrained SSM  have the shortest Euclidean distance from those of the ground truth system. Moreover, the better estimate of the eigenvalues of the system dynamics can be correlated with better open-loop performance, as reported in Table~\ref{fig:OL_MSE}.
Additionally, $\texttt{S-RNN}$ is the only model learning complex eigenvalues. This  can be further examined through the  physical interpretation of the eigenvalues given  as follows: real parts represent gains of the system, while 
the imaginary parts define the frequencies of the dynamics signals.
Hence,   $\texttt{S-RNN}$ model learned to associate the periodic behavior of the training data with the main system dynamics given by $\tilde{{\bf A}}$ transition matrix.
However, this is not correct association because the periodicity of the training data is the consequence of the periodic nature of the control inputs $\mathcal{A}$, $\mathcal{B}$  and disturbance trajectories $\mathcal{D}$ (day and night patterns).
   This may provide an explanation of why $\texttt{ODE}$  models outperform $\texttt{S-RNN}$ in the open-loop prediction. 
  Moreover, it can explain a remarkable capability of the  $\texttt{ODE}$  models in also predicting the unobserved states trajectories, despite not being explicitly trained to do so.
  In contrast,  $\texttt{S-RNN}$ fails to get even close to true trajectories of the unobserved states, as shown in Figure~\ref{fig:rnntrace}.
\begin{table}[htb]
    \centering
    \caption{Comparison of Eigenvalues for $\tilde{{\bf A}}$ transition matrix}
    \label{tab:eig}
\begin{tabular}{lllll}
\toprule
{} &               $\lambda_1$ &                 $\lambda_2$ &                                 $\lambda_3$ &                                $\lambda_4$ \\
\midrule
True     &   1.0 & 0.99 &  0.98& 0.25  \\
$\texttt{S-RNN}$    &   0.99 &  0.11+0.11i &  0.11-0.11i &-0.05  \\
$\texttt{ODE}_B$  &  1.0 &  0.88 & 0.21 &   0.02                                         \\
$\texttt{ODE}_G$    &   1.0 &     0.62 &                    0.15 &                   -0.01 \\
$\texttt{ODE}_W$    &1.0  &                     0.76 &    0.47  &                   0.07   \\
$\texttt{cODE}_B$   &   1.0 &     0.89 &                    0.15 &                   -0.03 \\
$\texttt{cODE}_G$ &    1.0 &     0.60 &                    0.21 &                   0.02 \\
$\texttt{cODE}_W$ &   1.0 &     0.65 &                    0.25 &                   0.03 \\
\bottomrule
\end{tabular}
\end{table}

 \subsection{Comparison of the Learned State Transition Parameters}
\begin{figure*}[htb]
     \centering
     \includegraphics{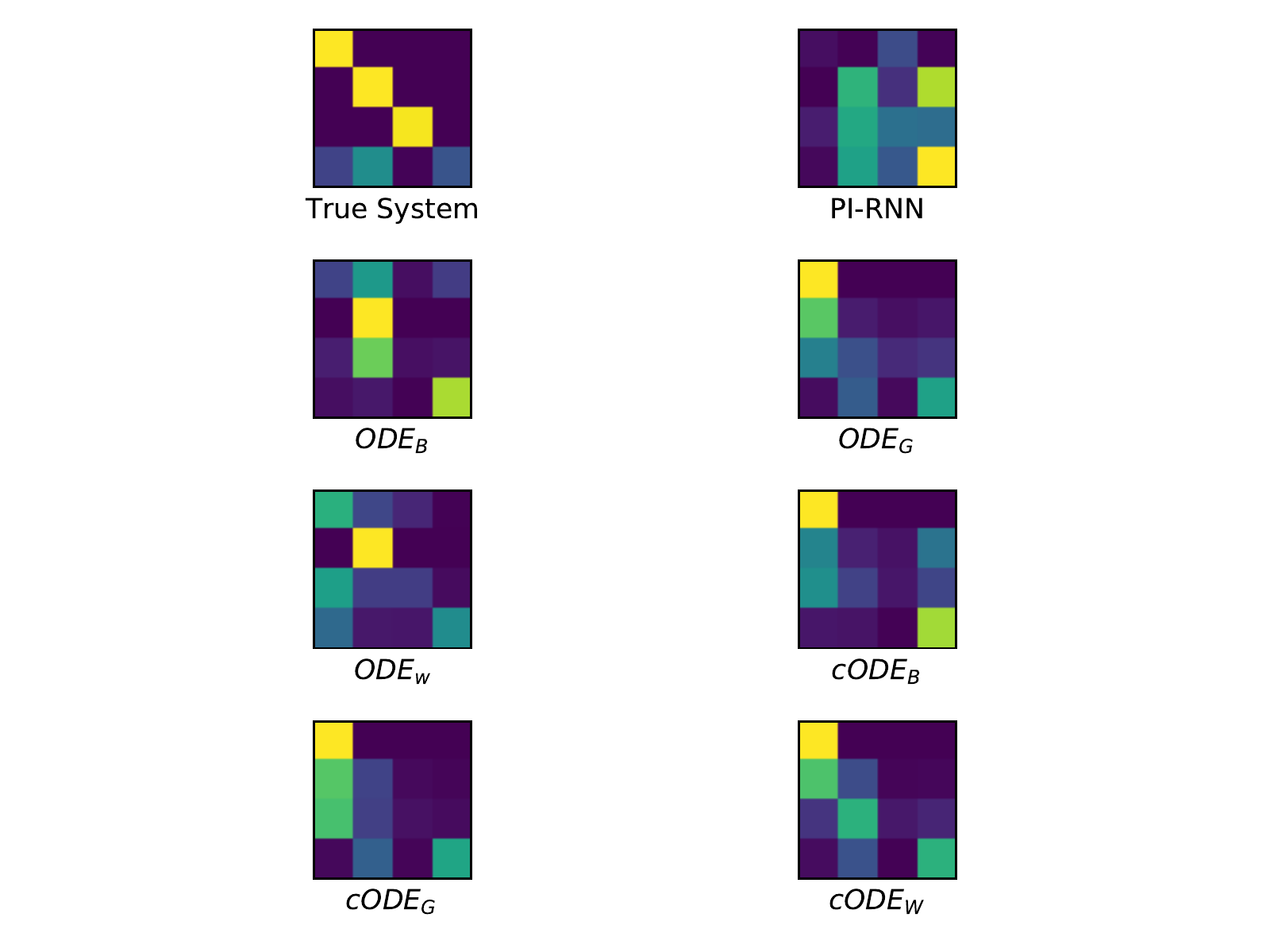}
     \caption{Heat maps of learned ${\bf \tilde{A}}$ matrices}
     \label{fig:heatmaps_A}
 \end{figure*}
 Figure~\ref{fig:heatmaps_A} compares the heat maps of the state transition matrix parameters of the learned models ${\bf \tilde{A}}$ with  ground truth values ${\bf {A}}$.
  The comparison of the learned model parameters of $\texttt{S-RNN}$ with $\texttt{ODE}$  models is less clear than in the case of eigenvalues.
  Nevertheless, we can spot that $\texttt{ODE}$ models are slightly sparser than $\texttt{S-RNN}$. 
Especially, $\texttt{ODE}_W$ models learn the most similar sparsity patterns in visual comparison with the diagonal structure of the ground truth model.
However, the $\texttt{ODE}_W$  is outperformed by $\texttt{ODE}_B$ model in the open-loop prediction task, indicating that not learning the true parameters but having the eigenvalues correct matters the most. 
Without any sparsity regularizations or structural priors on the  ${\bf \tilde{A}}$ matrix, no model can exactly identify the ground truth model parameters. However, this does not prevent the models to learn physically consistent open-loop dynamical trajectories with large time horizons. This might suggest that the solution to this system identification problem is not unique. In conclusion, it is important to say that a more rigorous analysis of the learned system parameters and eigenvalues, as well as the model structure, needs to be made to verify or falsify the qualitative statements in this appendix.

\end{document}